\documentclass[nofootinbib,twocolumn,reprint,aps,superscriptaddress, floatfix]{revtex4}

\usepackage{amsmath, amssymb}
\usepackage[T1]{fontenc} 
\usepackage{graphicx}
\usepackage{natbib}

\usepackage{mathrsfs}
\usepackage{mathtools}

\usepackage{color}

\newcommand{\eq}[1]{(\ref{#1})}

\newcommand{\be}{\begin{eqnarray}}
\newcommand{\ee}{\end{eqnarray}}

\usepackage[
colorlinks=true,        % color link
citecolor=blue,         % cite color
linkcolor=blue,         % link color
urlcolor=blue           % url color
]{hyperref}             % create hyperlinks
\usepackage{xcolor}    

\begin{document}

\title{A Unified Dark-Matter--Driven Relativistic Bondi Route to Black-Hole Growth from Stellar to Supermassive Scales}

\author{Chian-Shu Chen}
\affiliation{Department of Physics, Tamkang University, New Taipei 251, Taiwan}
\affiliation{Physics Division, National Center for Theoretical Sciences, Taipei 10617, Taiwan}
\email{chianshu@gmail.com}
\author{Feng-Li Lin,}
\affiliation{Department of Physics, National Taiwan Normal University, Taipei 11677, Taiwan}
\email{fengli.lin@gmail.com}

\date{\today}

\begin{abstract}
Observations of luminous quasars at $z\gtrsim7$ reveal supermassive black holes (SMBHs) with inferred masses $M_{\rm BH}\sim10^9 \, M_\odot$ formed within the first $\sim700$~Myr of cosmic history. Standard growth channels \textrm{---} Eddington-limited gas accretion and hierarchical mergers \textrm{---} face severe timescale restrictions. We consider a super-Eddington accretion mechanism aided by the Bondi accretion of a minimal model of self-interacting dark matter (SIDM).
%with the schematic accretion rate $\dot{M}=\Gamma M + K M^2$. 
We demonstrate that in a {\it critical regime} with a near-relativistic sound speed, the Bondi accretion yields an accretion rate that depends only on the mass $m$ of SIDM, thus it is universal to the ambient environment.
%that is, the Bondi coefficient $K$ is determined entirely by the mass $m$ of microscopic dark matter, imprinting the microscopic DM mass directly onto the black-hole growth function.
%Due to this universal feature, the SIDM-Bondi-aided Eddington accretion has a closed-form solution and predicts a microphysics-controlled transition mass $M_\star =\Gamma/K$ at which relativistic Bondi growth dominates. 
This critical accretion mechanism for $m\gtrsim 10^{-2}\; {\rm eV}$ can grow seeds as small as $10\,M_\odot$ primordial black holes (PBH) in the early Universe into $10^9$ \textrm{--} $10^{10}\,M_\odot$ SMBHs by $z\sim7$ without fine-tuned environments. Therefore, given a mass distribution of PBHs and a value of $m$, the mass function of primary black holes at late time can be fully determined with masses ranging from stellar to SMBHs. This connects the microscopic physics of dark matter to astrophysical observations of black holes.  
\end{abstract}

\maketitle

\section{Introduction}
%\medskip
%\noindent\textbf{Introduction.—}
The discovery of luminous quasars hosting supermassive black holes (SMBHs) with masses $\sim10^{9}M_\odot$ at
$z\gtrsim6-7$~\cite{Banados:2017unc,Yang2020Poniuaena,Larson2023CEERS,Larson2023,Kokorev2023UNCOVER,Maiolino2024,Furtak2023JWST,Harikane2023_JWST_AGN,Kocevski2024,Kovacs:2024zfh,Bogdan:2023ilu,Juodzbalis2024,Kocevski2025,Taylor2025}
%,Wang2021,Natarajan:2023rxq,Regan:2024wsu,Kovacs:2024zfh,Guia:2024toq,2025ApJ...986..165T,2025ApJ...986..126K,Matthee:2023utn}
presents one of the sharpest timing challenges in cosmology:
assembling such masses within the first $\sim 700$\,Myr strains the limits of
standard formation channels based on Eddington-limited gas accretion,
direct-collapse seeds, or hierarchical mergers
\cite{Wise2019_MassiveBH_preGalactic,Bromm2002FirstStars,Oh2002SecondGen,Begelman2006_DCBH,Volonteri2010,Inayoshi2020,Smith2019SMBHReview,Greene2020,Prole:2023toz}.  
Pop~III remnants require prolonged, uninterrupted Eddington phases, while forming black holes from direct collapse occurs only in rare metal-poor environments.
Merger-driven pathways --- from classical hierarchical assembly
\cite{Begelman1980_MHBBinaries,Volonteri2003} to modern cluster-driven and gravitational-wave-era hierarchical merger models
\cite{MillerHamilton2002,PhysRevD.100.043027,GerosaFishbach2021} --- can build stellar or 
intermediate-mass black holes (IMBHs), but generally lack the duty cycle and retention
efficiency needed to reach $\sim10^9M_\odot$ by $z\sim7$
\cite{Inayoshi2020,Greene2020}.  
These difficulties motivate additional, non-baryonic growth channels.

A natural channel beyond the Eddington accretion is aided by the Bondi accretion, with the accretion rate schematically given by
\be\label{master_acc}
\dot{M}=\dot{M}_{\rm Edd} + \dot{M}_{\rm Bondi} = \Gamma M + K M^2
\ee
where $M$ is the mass of the central black hole, $\Gamma = 4\pi G_N m_p/\eta c\sigma_T\approx4.5\times10^{-8}{\rm yr^{-1}}$ (assuming radiation efficiency $\eta\simeq 0.1$) is the baryonic Eddington coefficient, and $K$ is the coefficient for Bondi accretion. It can be determined by the equation of state (EoS) of the underlying fluid and will, in general, depend on the sound speed $a$ and the ambient mass density $\rho_0$, e.g., $K\sim a^{-3} \rho_0$ for the fluid of pressureless matter such as cold dark matter (CDM).  Since dark matter (DM) prevails in the Universe and forms the main part of the typical halos, it should be responsible for feeding the central black hole in a halo via Bondi accretion. The Bondi accretion is sub-dominant when $M$ is small, and will start to dominate over the Eddington accretion at the transition mass $M_\star \simeq {\Gamma}/K$. 

For the Bondi-aided Eddington accretion to achieve the formation of $10^9 M_\odot$ SMBHs at $z\simeq 7$, it requires to have small $M_\star$, thus large $K$, around $z\sim 20-30$, during which the galaxy-scale halos with masses $\sim 10^{11}-10^{12} M_\odot$ forms, so that the central black hole can acquire most of the halo's mass to reach $10^9 M_\odot$ at $z\simeq 7$ via the dominant Bondi accretion. However, this condition will usually be destroyed by a non-negligible thermal sound speed, yielding a small Bondi radius, so that the Bondi accretion will not be complete. The above no-go results happen for CDM \footnote{See the supplementary material for more detailed discussions}, unless a heavy central core, such as an IMBH, appears around $z\sim 20-30$, which requires specific scenarios \cite{Begelman2006_DCBH, Chiu2025PRL}. This then motivates us to consider the alternative models of DM other than CDM, and the associated Bondi accretion. 

A feature of CDM is almost pressureless; it is then reasonable to lift it by introducing the self-interactions, denoted as self-interacting DM (SIDM). This model was introduced to resolve some small-scale issues, e.g., the core–cusp, too-big-to-fail, and missing satellites problems arising from the $\Lambda$CDM, by requiring the following constraints on the SIDM's scattering cross-section $\sigma$ to mass $m$ ratio \cite{SpergelSteinhardt2000, TulinYu2018} from merging clusters \cite{Clowe2006,Randall2008},
\be\label{sigma_over_m}
{\sigma \over m} \sim0.1-1~\rm{cm^2 g^{-1}}\;.
\ee
This simple constraint cannot fully pin down a SIDM model, leaving many possibilities open. Here, we shall mention the pioneer works by Ostriker~\cite{Ostriker:1998fa, Ostriker:1999ee} by introducing the SIDM with velocity-dependent cross-section constrained by \eq{sigma_over_m} to explain the formation of SMBHs via Bondi accretion and predict the Magorrian relation ($M_{\rm BH} \propto M_{\rm halo}$)~\cite{Magorrian:1997hw}. This approach is based on the particle picture of SIDM scattered by the Newtonian potential around a central black hole. The particle picture is intuitive and flexible in modeling the velocity dependence of the cross-section. In contrast, in this Letter, we will consider the relativistic perfect-fluid picture of a minimal SIDM that forms SMBHs via Bondi accretion. The fluid picture of SIDM captures the collective behavior of SIDM beyond the particle picture and explains the formation of SMBHs through a mechanism quite different from that in the particle picture.

From the particle physics point of view, the simplest SIDM is a relativistic scalar field $\phi$ of mass $m$ with polynomial self-interaction,
\begin{equation}\label{phi_n}
\mathcal{L}=\tfrac12(\partial\phi)^2-\tfrac12 m^2\phi^2-\tfrac{\lambda}{n!}\phi^n\;.
\end{equation}
In this Letter, we will show that there is a {\it critical regime} in the parameter space of $m$ and $\lambda$, in which the Bondi accretion rate depends only on $(m,\lambda)$, i.e., $K$ is just a function of $m$ after imposing the constraint \eq{sigma_over_m}, thus it is universal to all halo environments. This is different from the consideration in \cite{Feng2025DarkBondi}, where the Bondi accretion of SIDM is not in the critical regime.

The universality of the Bondi accretion rate leads to a few important results: (i) it will bypass the difficulties of CDM in forming the $10^9 M_\odot$ SMBHs at $z\simeq 7$ without introducing the specific halo environments, such as a heavy core;  (ii) given a SIDM model like \eq{phi_n}, the necessity of achieving successful Bondi-aided super-Eddington accretion will fully determine the model parameters, along with \eq{sigma_over_m}; (iii) Once (ii) is done, we can obtain the growth curve of the central black hole, and also the mass function of the primary black holes by assuming a initial mass function for primordial black holes (PBHs).  These results unify the microscopic particle physics of DM with the astrophysical observations of SMBHs, and simultaneously provide the mass function of the PBHs, which will be crucial to estimate the event rates of their mergers detected by gravitational wave observations. 

Despite the success of our model in explaining the SMBHs at $z\simeq 7$, our SIDM model yields a velocity-independent cross section $\sigma_{\phi^4}={\lambda^2 \over 64\pi m^2}$ and cannot help explain the multiscale observations, known as the Goldilocks problem of dark matter halos, see e.g.,  \cite{TulinYu2018}: galaxy-scale constraints require higher values of $\sigma/m$, but the cluster scale constraints require lower values. This implies that we need to add a new ingredient to our SIDM model to introduce velocity dependence of the cross-section to resolve the Goldilocks problem. One way is to introduce another SIDM with a velocity-dependent cross-section to solve the Goldilocks issue, but it will not contribute significantly to Bondi accretion, as is usually expected. Another way is to introduce a mediator field of mass $m_I$ to our $\phi^4$-SIDM, which can solve the Goldilocks problem as shown in, for example,  \cite{Tsai:2020vpi} with $m_I \gtrsim {\cal O} ({\rm MeV})$. On the other hand, as argued in \textit{supplementary}, a heavy mediator field will not change the nature of the fluid of $\phi^4$-SIDM for considering the critical Bondi accretion. This can be simply seen by integrating out the heavy mediator field to restore the $\phi^4$-SIDM with a renormalized quartic coupling.

\section{$\phi^4$-SIDM and critical Bondi accretion}

%\medskip
%\noindent\textbf{$\phi^4$-SIDM and critical Bondi accretion.—} 
To be specific, we will consider the $n=4$ case of \eq{phi_n}, denoted as $\phi^4$-SIDM \footnote{In supplementary material, we demonstrate that the critical regimes for the universal Bondi accretion rate also occur for all $n=4,5,\cdots$. The $n=4$ gives the most critical behavior, and it is also the simplest and most natural model. Thus, we consider $n=4$ in detail.}.  In \cite{Colpi:1986ye, Zhang:2023hxd}, it is shown that under the isotropic self-gravitating regime, which is just the same regime for considering Bondi accretion, the $\phi^4$-SIDM forms a fluid with the following EoS relating energy density $\rho$ and pressure $p$, 
\begin{equation}
\frac{\rho}{\rho_B}=3\frac{p}{\rho_B}+4\sqrt{\frac{p}{\rho_B}},
\label{eos_main}
\end{equation}
where the characteristic density scale $\rho_B=\frac{3 m^4}{2\lambda}$.  The narrow window of \eq{sigma_over_m} gives a tight relation between $\lambda$ and $m$, and its median value can be converted into 
\begin{equation}
\rho_B = 6.96\times10^{18}\Big(\frac{m}{\rm{GeV}}\Big)^{5/2}~\rm{kg~m^{-3}}\;.
\label{rhoB_scaling_main}
\end{equation}

With the EoS \eq{eos_main} and adopting the relativistic Euler equation \cite{Shapiro:1983du}, we can obtain the Bondi accretion rate of $\phi^4$-SIDM ($c=G_N=1$) \footnote{To arrive \eq{eq:accretion-SIDM}, we need to use the relation between $\rho_{0,s}$ and $\rho_{\rm DM}$. The former is the mass density at the sonic horizon, and the latter is the asymptotic mass density. The detailed derivations of the following results on Bondi accretion of $\phi^4$-SIDM can be found in the supplementary material. In this work, we adopt the units with $c=G_N=1$, where $c$ is the speed of light, and $G_N$ the Newton constant.}, 
\begin{equation}
\dot{M}_{\rm Bondi}
=2\pi\frac{(1 + 3 a_s^2)^3}{a_s^3(9a_s^2-1)}\,M^2\,\rho_{\rm DM},
\label{eq:accretion-SIDM}
\end{equation}
where $a_s$ is the sound speed $a:=\sqrt{\partial p\over \partial \rho}$ at the sonic horizon, and the $\rho_{\rm DM}$ is the ambient DM mass density. This Bondi accretion rate seems divergent as $a_s \rightarrow 1/3$. It is quite different from the Bondi accretion rate $\sim a^{-3}_s M^2 \rho_{\rm DM}$ for CDM, which cannot be divergent as $a_s$ is non-negligible. Moreover, as argued in \textit{supplementary} and discussed right below, this seemingly divergence of the accretion rate at the critical limit $a_s \rightarrow 1/3$ turns out to be physically finite and environment-independent. This is the key result to be adopted for the SMBH formations.

As our goal is to achieve the dominant Bondi accretion, we consider only the critical regime by introducing $\epsilon:=a_s^2 - 1/9 \ll 1$. In this critical regime, the consistency relations of Bondi accretion formalism yield\footnote{For details, see \textit{supplementary}. There, we first obtain $\epsilon\simeq {2\over 9} {\rho_{\infty} \over \rho_B}$, where $\rho_{\infty}$ the asymptotic energy density and generally differs from the mass density $\rho_{\rm DM}$ unless the asymptotic pressure $p_{\infty}$ is vanishing. We further pointed out that $p_{\infty}\simeq {81\over 64} \rho_B \epsilon^2$ is one order higher in $\epsilon$ than $\rho_{\infty} \simeq {9\over 2} \rho_B \epsilon$. We can then treat $p_{\infty} \simeq 0$, so that $\rho_{\infty}\simeq \rho_{\rm DM}$. }
\be\label{eps_rel}
\epsilon \simeq {2\over 9} {\rho_{\rm DM} \over \rho_B} \simeq {16 \over 9} a_{\infty}^2\;,
\ee
where $a_{\infty}$ is the asymptotic sound speed.
By \eq{eps_rel}, \eq{eq:accretion-SIDM} is reduced to 
\be\label{Bondi_crit}
\dot{M}_{{\rm Bondi}}
= 64\pi\,\rho_B\,M^2 \;,
%\\ &=&4.59 \times 10^{-10} \; \Big({M \over M_\odot}\Big)^2 \; \Big(\frac{m}{\rm{GeV}}\Big)^{5/2} \; {M_\odot \over {\rm yr}}\;, \quad 
\ee
yielding a universal coefficient of Bondi accretion to all halo environments, i.e, $\rho_{\rm DM}$ independent, 
\be\label{univ_K}
K=64 \pi^2 \rho_B \propto m^{5/2}\;.
\ee
By \eq{rhoB_scaling_main}, $K$ is fully determined by the scalar mass $m$. This relates the microscopic DM physics to the Bondi accretion rate in the formation of SMBHs. 

Furthermore, using both equalites of \eq{eps_rel} yields the Bondi radius
\be\label{r_feed}
r_B:={2 M \over a_{\infty}^2} \simeq {81\over 16} {\rho_B \over \rho_{\rm DM}} M\;.
\ee
The Bondi radius is the effective feeding radius to capture the DM into the accretion flow. In contrast to the case of CDM, in which the $a_{\infty}$ is non-negligible, the Bondi radius for $\phi^4$-SIDM can be very large even for IMBHs. Below, we will explicitly demonstrate that the universal accretion rate \eq{Bondi_crit} and large feeding radius \eq{r_feed} are crucial to reach $10^9 M_\odot$ at $z\simeq 7$ by appropriately choosing the parameter $m$ of the SIDM mass.

\section{Unified growth equation and transition mass}
%\medskip
%\noindent\textbf{Unified growth equation and transition mass.—}
The universal Bondi coefficient $K$ in \eq{univ_K} is environment-independent, and also time-independent. Thus, the Bondi-aided Eddington accretion \eq{master_acc} admits a simple and analytic solution,  
%Including baryonic Eddington inflow with coefficient  
%$\Gamma=4\pi G m_p/\eta c\sigma_T$ (the radiation efficiency $\eta=0.1$), the full mass evolution 
%becomes
%\begin{equation}
%\dot M=\Gamma M+K(m) M^2.
%\end{equation}
%This Riccati equation 
%admits a closed analytic solution:
\begin{equation}
M(z)=\frac{M_0\,e^{\Gamma \tau(z)}}
{\,1-\dfrac{K}{\Gamma}M_0\!\left(e^{\Gamma \tau(z)}-1\right)}
\label{eq:MassGrowthClosedFormMain}
\end{equation}
where $M_0$ is the initial seed mass, and the conversion between cosmic time and redshift follows the standard relation $\tau(z) = \int_{z}^{z_i}\frac{dz'}{(1+z')\,H(z')}$ where $H(z)=H_0\sqrt{\Omega_r(1+z)^4+\Omega_m(1+z)^3+\Omega_\Lambda}$. All time–redshift conversions use Planck 2018 $\Lambda$CDM cosmology ($H_0=67.4$, $\Omega_m=0.315$, $\Omega_\Lambda=0.685$, and $\Omega_\gamma=9.2\times10^{-5}$)~\cite{Planck2018}. Equation~\eqref{eq:MassGrowthClosedFormMain} is the central analytic result of this
Letter. Once the SIDM mass $m$ is chosen, the coefficient $K \!\propto\!m^{5/2}$ uniquely determines the entire growth history, independent of halo structure or baryonic microphysics.

%the entire evolutionary history is fixed once $m$ is specified through  
Although the relativistic SIDM Bondi rate is universal, the total mass a PBH can accrete is ultimately limited by the finite SIDM reservoir of its host halo \footnote{In the supplementary material, with the chosen mass parameter $m$ (thus $\rho_B$) for the Bondi accretion, we show that the corresponding EoS can yield stable TOV configurations of the SIDM halos with their total mass (size) four to five orders larger (smaller) than the ones of the typical galaxy-type halos. This alternatively demonstrates that the $\phi^4$-SIDM halos can have enough masses to feed the SMBHs larger than $10^9 M_\odot$.}. We adopt a conservative supply model in which the ambient DM density around the galaxy halos, 
\be\label{rho_DM_z}
\rho_{\rm DM}(z)=5\times10^{-22}(1+z)^3 \; \Delta_{\rm{vir}} \; {\rm kg\,m^{-3}} 
\ee
with a typical virial overdensity $\Delta_{\rm{vir}}\simeq300$ movitavted by spherical-collapse modeling and simulations~\cite{BryanNorman1998,WhiteRees1978,Blumenthal1984,Bryan2013EAGLE}. In this environment, the Bondi radius given by \eq{r_feed} grows approximately linearly with black-hole mass and decreases with redshift, 
\begin{equation}\label{r_B_z}
    r_{B}(z,m)\simeq3.55\times10^2 \; \frac{(m/\rm{eV})^{5/2}}{(1+z)^3}\frac{M(z)}{M_{\odot}}~~\rm{pc}\;,
\end{equation}
so that at late times $r_{B}$ can engulf an entire SIDM core and the evolution becomes supply-limited rather than Bondi-limited. The competition between the Eddington term and relativistic SIDM Bondi inflow is encoded in the transition mass
$M_\star=\Gamma/K \propto m^{-5/2}$, which depends steeply on the SIDM particle mass. The numerical connection between $m$ and $M_\odot$ is summarized in Table \ref{Table_I}. Note that  $M_\star \gtrsim10^{9}M_{\odot}$ if $m$ is near the lower bound of \eq{m_bound}, so PBHs remain Eddington-limited.

\begin{table}[t]
\centering
\begin{tabular}{c c c c}
\hline\hline
$\;m\;({\rm eV})$  &  $M_\star \;(M_\odot)$ &  $z_\star$ & Growth regime \\
\hline
$10^{-3}$ & $1.5307\times10^{9}$ & none & Bondi never dominates \\
$10^{-2}$ & $4.8403\times10^{6}$ & 7.8 & Bondi dominates around $z\sim11$ \\
$10^{-1}$ & $1.5307\times10^{4}$ & 11.31 & Early Bondi onset \\
$1$       & $4.8403\times10^{1}$ & 22.83 & Bondi dominates almost immediately \\
\hline\hline
\end{tabular}
\caption{Transition mass $M_\star=\Gamma/K$ at which relativistic SIDM Bondi inflow overtakes baryonic Eddington growth, and we denote $z_\star$ as the corresponding cosmic redshift marking the onset of this Bondi-dominated accretion phase.  Because $K \!\propto\!m^{5/2}$, the transition scale spans nearly eight orders of magnitude across the range $m=10^{-3}$–$1$\,eV. For light SIDM, $M_\star\!\gtrsim\!10^{9}M_\odot$ so PBHs remain Eddington-limited, whereas for $m\!\gtrsim\!10^{-2}$\,eV the Bondi phase 
switches on early, imprinting a strong microphysical signature on the black-hole mass function.
} \label{Table_I}
\end{table}

\begin{figure}[t]
\centering
  \includegraphics[width=0.48\textwidth]{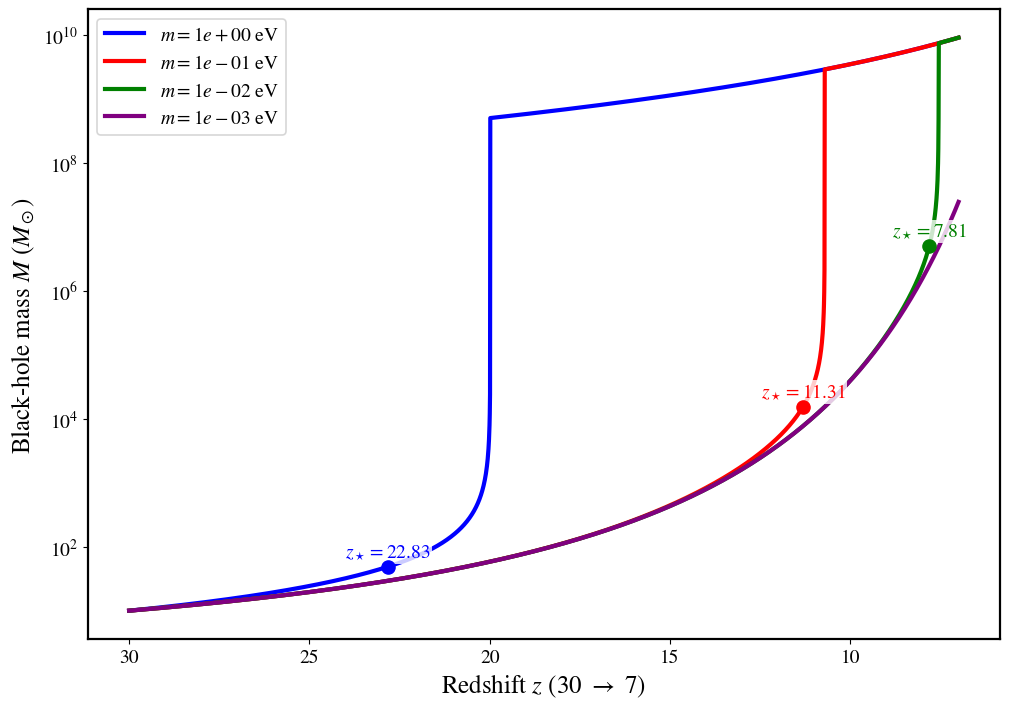}
\caption{
Mass evolution of a $10\,M_\odot$ seeding black hole from $z=30$ to $z=7$, including Eddington accretion and relativistic SIDM Bondi inflow.
Shown are four representative SIDM masses, $m=1,~10^{-1}, 10^{-2}$, and $10^{-3}$ eV, evolved in a Press--Schechter median halo enhanced by an order of magnitude
with a total SIDM supply fraction $f_{\rm halo}=0.9$. The heavier SIDM reaches the transition mass $M_\star$ at the corresponding redshift $z_\star$, marked by crosses) earlier and experiences a rapid growth phase until saturating at the finite halo supply. These results demonstrate that the accretion process involves three stages: (i) a long early-time Eddington-dominated phase; (ii) supply-limited Bondi-dominated phase starting around $z=z_\star$ and ending at $z=z_B$ after exhausting the SIDM of the inner halo core, indicated by the sharp turns into a universal plateau; (iii) late-time Eddington-dominated phase, i.e., shown by the plateau.
}
\label{fig:PBHGrowth}
\end{figure}
\begin{figure}[t]
\centering
\includegraphics[width=0.95\linewidth]{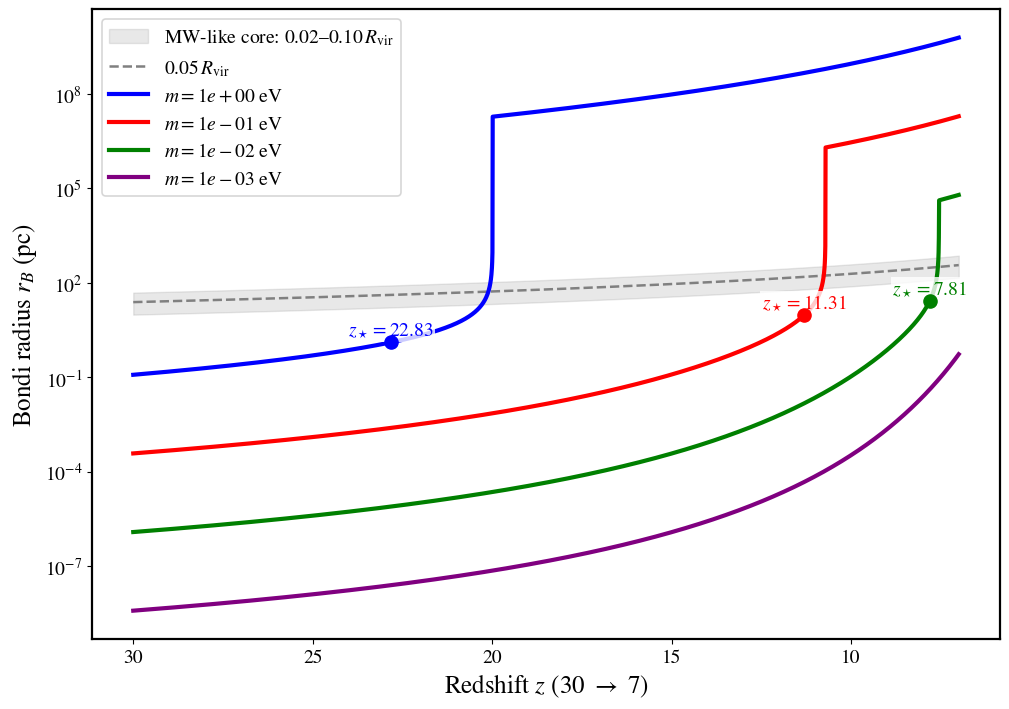}
\caption{
Counterpart evolution of the Bondi radius $r_B(z)$ of Fig. \ref{fig:PBHGrowth}, assuming the average-halo density of \eq{rho_DM_z}. The $y$-axis is in parsecs (logarithmic), while the $x$-axis is linear in redshift from $z=30$ (left) to $z=7$. This demonstrates how the Bondi radius expands across cosmic epochs, eventually approaching or crossing the Milky Way-like inner-halo scale ($r_{\rm{core}}\approx0.05R_{\rm{vir}}$, represented by the gray dashed line and shaded band). This enables efficient SIDM feeding in the late growth phase to form SMBHs with masses $\geq 10^9 M_\odot$ if the crossing occurs at $z > 7$.
}
\label{fig:rB(z)}
\end{figure}

Given $\rho_{\rm DM}$ of \eq{rho_DM_z}, the crticial parameter in the first equality of \eq{eps_rel} can be expressed explicitly,   
\be
\epsilon(z,m) = 1.52 \times 10^{-16} {(1+z)^3 \over (m/{\rm eV})^{5/2}} \;.  
\ee
To ensure the $\phi^4$-SIDM Bondi accretion is in the critical regime, we require $\epsilon \lesssim 10^{-3}$ at $z\simeq 7$. This then imposes a lower bound on $m$, i.e.,
\be\label{m_bound}
m\gtrsim 10^{-3.34} {\rm eV}\;.
\ee
Choosing some representative values of $m$ satisfying \eq{m_bound}, we exemplify the mass evolution of \eq{eq:MassGrowthClosedFormMain} in Fig. \ref{fig:PBHGrowth} for an initial mass of $10 M_\odot$, and the Bondi radius counterpart $r_B$ of \eq{r_B_z} in Fig. \ref{fig:rB(z)}. As expected, the heavier SIDM has a sharper growth and a lower transition mass $M_\star$ (marked by crosses in Fig. \ref{fig:PBHGrowth}), so that it is easier to yield the $10^9 M_\odot$ SMBHs. In arriving Fig. \ref{fig:PBHGrowth} and \ref{fig:rB(z)}, we assume that most of the SIDM resides in the inner halo core, so that the Bondi accretion is supply-limited once the inner core SIDM is accreted into the center black hole. Specifically, the Bondi-dominated phase starts at $z=z_\star$, where $M=M_\star$, and soon can engulf most of the inner core SIDM's mass into the black hole not long after the Bondi radius $r_B$ exceeds $r_{\rm core}$, the inner core size of the halo. After consuming most of the SIDM in the inner core, the accretion process abruptly exits the Bondi-dominated phase at $z=z_B$ and re-enters the universal Eddington-dominated phase. These transitions are indicated in Fig. \ref{fig:PBHGrowth} and \ref{fig:rB(z)} by the sharp turns into an Eddington plateau at $z=z_B$. To maintain physical consistency, total growth is constrained by a mass-supply cap, which we assume to be $f_{\rm halo}=0.9$ fraction of the total halo mass, thereby restricting the final BH mass to about $90\%$ of the host halo's total mass budget around $z=z_B$.

%\subsection{QCD PBH mass function at formation}

%As exemplified in Fig. \ref{fig:PBHGrowth}, light SIDM ($m\sim10^{-3} \textrm{--} 10^{-4}$~eV) yields $M_\star \gtrsim10^{9}M_{\odot}$, keeping black holes Eddington-regulated throughout cosmic history. Heavier SIDM ($m\sim10^{-2}-1$~eV) drives $M_\star$ down to $10^{1-6}M_{\odot}$, enabling an early Bondi-dominated phase that rapidly amplifies mass until limited by halo supply. This sharp scaling of $M_\star$ as summarized in Table \ref{Table_I} provides an observationally testable microphysical imprint on the evolving black-hole mass function.

Due to the window of $\sigma/m$ given in \eq{sigma_over_m}, the SIDM develops constant-density cores through heat conduction driven by DM self-scattering, with simulations typically giving $r_{\rm core} \sim 0.02\textrm{--}0.10 R_{\rm vir}$ \cite{Balberg:2002ue,Kaplinghat:2015aga,TulinYu2018}. This environment is modeled as a Milky Way-like dark matter core nested inside a massive host halo. We adopt a halo reference mass of $M_{\rm{ref}}\approx5\times10^{9}M_{\odot}$ at $z=10$, derived from the Press-Schechter (PS) formalism to represent high-density fluctuation peaks from $z=30$ to $7$. Here $R_{\rm{vir}}$ is the virial radius enclosing an overdensity factor $\Delta_{\rm{vir}}\sim200 \textrm{--} 300$ relative to the mean matter density, providing a natural scale for the halo boundary. The characteristic inner core size of the host halo is modeled after a Milky Way-like profile (gray dashed line in Fig.~\ref{fig:rB(z)}), serving as a structural benchmark for the onset of supply-limited accretion. We overlay this core band to indicate when the Bondi radius $r_{B}(z)$ exceeds the SIDM core~in Fig.~\ref{fig:rB(z)}, signaling the onset of supply-limited growth. This is the key ingredient for guaranteeing the formation of SMBHs with masses comparable to the total mass of the hosting dark halos.  

Our results in Fig. \ref{fig:PBHGrowth} and \ref{fig:rB(z)} indicate that the formation of SMBHs can end around $z=z_B \simeq 25-10$ with their mass about $f_{\rm halo} M_{\rm halo}(z\simeq z_B)$, earlier than the observed time around $z=7$ by the observations of the James Webb Space Telescope (JWST). The subsequent cosmic evolution of halo mass assembly through hierarchical mergers will increase halo mass but not the mass of the central SMBH. Thus, the halo mass typically scales as $(1+z)^{- \alpha}$ with $\alpha=3-4$ \cite{o2015probing, griffen2016caterpillar, lacey1994merger, Bond:1990iw}. This will then reduce the mass ratio $M_{\rm BH}(z_B)/M_{\rm halo}(z)$ by a factor of $({1+z \over 1+z_B})^{\alpha}$. For $\alpha=3$, $z=7$ and $z_B=25-10$, we will obtain $M_{\rm SMBH}/M_{\rm halo} \simeq 0.3 - 0.03$. The order of this mass ratio is remarkably consistent with recent JWST observations of high-redshift quasars and galaxies \cite{Harikane2023_JWST_AGN, Maiolino2024, Bogdan:2023ilu}, but exceeds the local Magorrian relation ($\sim 10^{-3}$) \cite{Magorrian:1997hw, KormendyHo2013}. This indicates that the SMBHs formed in our "early blossom of SMBHs vs late-time assembly of halos" scenario are, in general, "overmassive" compared with their lower-redshift counterparts.

Comparing our super-Eddington channel to the other earlier proposals, the latter rely on environmental enhancements such as the catastrophic gas inflow in metal-poor halos triggered by bars-within-bars instabilities~\cite{Begelman2006_DCBH}, or the deep central potentials of fuzzy-DM solitons that pre-compress gas~\cite{Chiu2025PRL}. In contrast, our Bondi-aided SIDM proposal stems directly from the EoS due to the self-gravitating microscopic SIDM, yielding a universal relativistic Bondi coefficient $K$ and an analytic growth law that operates in ordinary halos, without requiring special astrophysical conditions, as long as a sufficient SIDM supply is available. This establishes a direct and falsifiable mapping between DM microphysics and the emergence of SMBHs.

\section{PBH seeds to mass function at $z=7$}
%\medskip
%\noindent\textbf{PBH seeds to mass function at $z=7$.—}
Our mass growth formula is universal and can be applied to any seeding black hole at all times. To exemplify, we consider the seeding black holes to be PBHs formed in the early Universe as the stellar-size seeds to provide the high-redshift quasars. In particular, the QCD phase transition briefly softens the EoS of the plasma, enhancing the collapse of horizon-scale perturbations 
and naturally producing PBHs with masses of order 
${\cal O}(1$--$10)\,M_\odot$~\cite{CarrHawking1974,Jedamzik1997,Byrnes2018}.  
These objects formed long before the first stars, avoiding all baryonic 
cooling requirements, and began their growth in a DM-dominated environment.  Within our framework, PBHs are therefore an ideal testbed 
for examining how the relativistic SIDM Bondi channel reshapes the mass function by $z=7$.

%We concentrate on the post-recombination era and earlier phases are deliberately omitted.  
%During the radiation-dominated era and the immediately post-equality 
%epoch, baryons remain tightly coupled to photons, DM densities are low, 
%and the Bondi radius of ${\cal O}(10)\,M_\odot$ PBHs is negligible; 
%accretion is therefore ineffective.  
%After recombination, baryons decouple from radiation and fall into DM 
%potentials, and SIDM densities increase sharply in halo centers.  
%Only from this stage onward does accretion become astrophysically 
%meaningful, and the analytic growth law obtained in Sec.~II can be applied self-consistently. 

%We take two benchmark DM mass, $m=10^{-2}$ eV and $10^{-6}$ eV, to illustrate the PBH mass growth from $z=30$ to $z=7$ for the initial PBH mass $M_0=10~M_{\odot}$. In Fig.~\ref{fig:PBHGrowth}, it shows that for $m=10^{-2}$ eV the Bondi accretion takes over as $M_{\rm{BH}}\sim10^{6}~M_{\odot}$, and the final mass is able to reach $\sim10^{10}~M_{\odot}$.  

%\medskip
%\noindent\textbf{BH mass function at $z=7$—}
To capture the range of QCD-era PBH formation scenarios, we consider three representative seed distributions, each normalized to $\int\psi_{0}(M_{0})dM_{0}=f_{\rm{PBH}}$: 

$(i)$ Log-normal\textrm{---} a sharply peaked spectrum near the horizon mass at $T\sim200$ MeV,
\begin{equation}
\psi^{\rm{LN}}_0(M_0)=\frac{f_{\rm PBH}}{\sqrt{2\pi}\sigma M_0}
\exp\!\left[-\frac{(\ln(M_0/M_c))^2}{2\sigma^2}\right],
\end{equation}
by choosing $M_{c}=5M_{\odot}$ and $\sigma=0.5$;  

$(ii)$ Power Law (scale-free tail)\textrm{---} capturing broad, extended PBH populations, 
\begin{equation}
\psi^{\rm{PL}}_0(M_0)=\frac{(1-\alpha) f_{\rm{PBH}}}{M^{1-\alpha}_{\rm{max}}-M^{1-\alpha}_{\rm{min}}} M^{-\alpha}_{0} \;,  
\end{equation}
by choosing $\alpha=2$, and $M_0\in [M_{\rm min}, M_{\rm max}]$ with  $M_{\rm{min}}=0.1M_{\odot}$ and  $M_{\rm{max}}=100M_{\odot}$.

$(iii)$ Critical-collapse\textrm{---} bearing a universal near-threshold scaling, 
\begin{equation}
\psi^{\rm{CC}}_0(M_0)=\tilde{A}M^{\frac{1}{\gamma}-1}_0\exp\Big[-\beta(M_{0})/M_{c})^{1/\gamma}\Big],
\end{equation}
by choosing $\gamma\simeq0.36$, $\beta\simeq1$, and $M_{c}=5M_{\odot}$. $\tilde{A}$ is fixed by normalization condition.

\begin{figure}[t]
\centering
\includegraphics[width=0.47\textwidth]{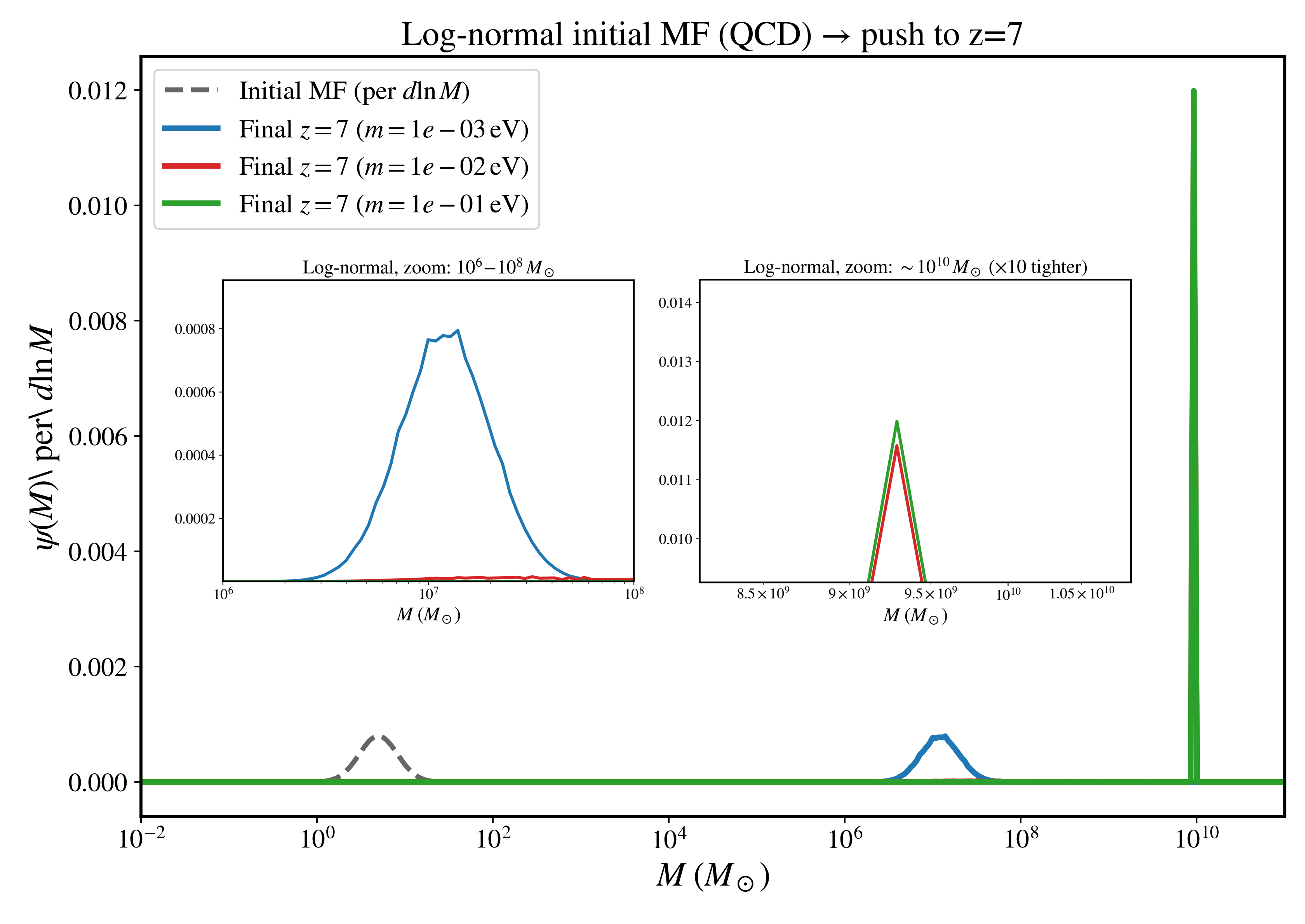}
\includegraphics[width=0.47\textwidth]{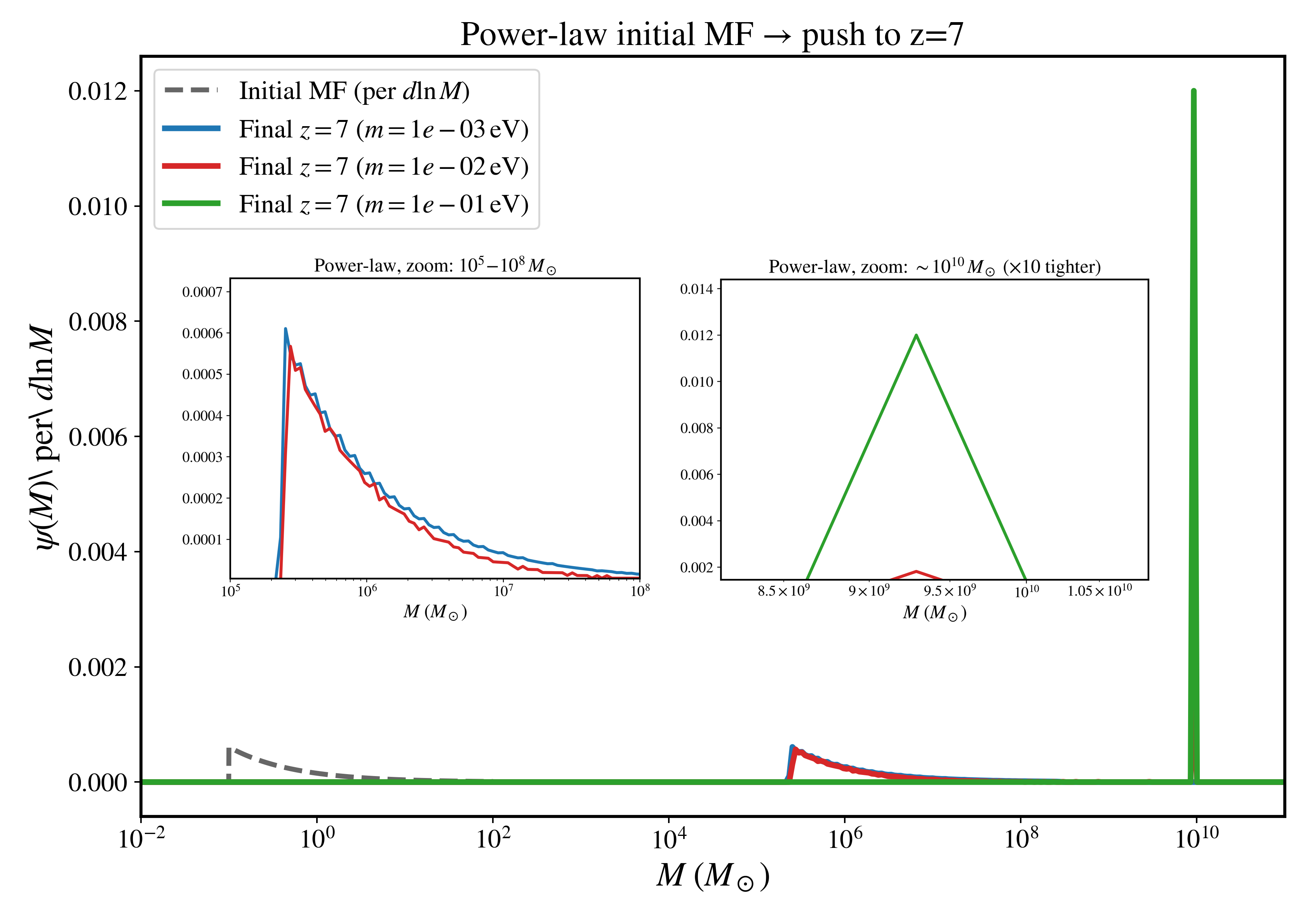}
\includegraphics[width=0.47\textwidth]{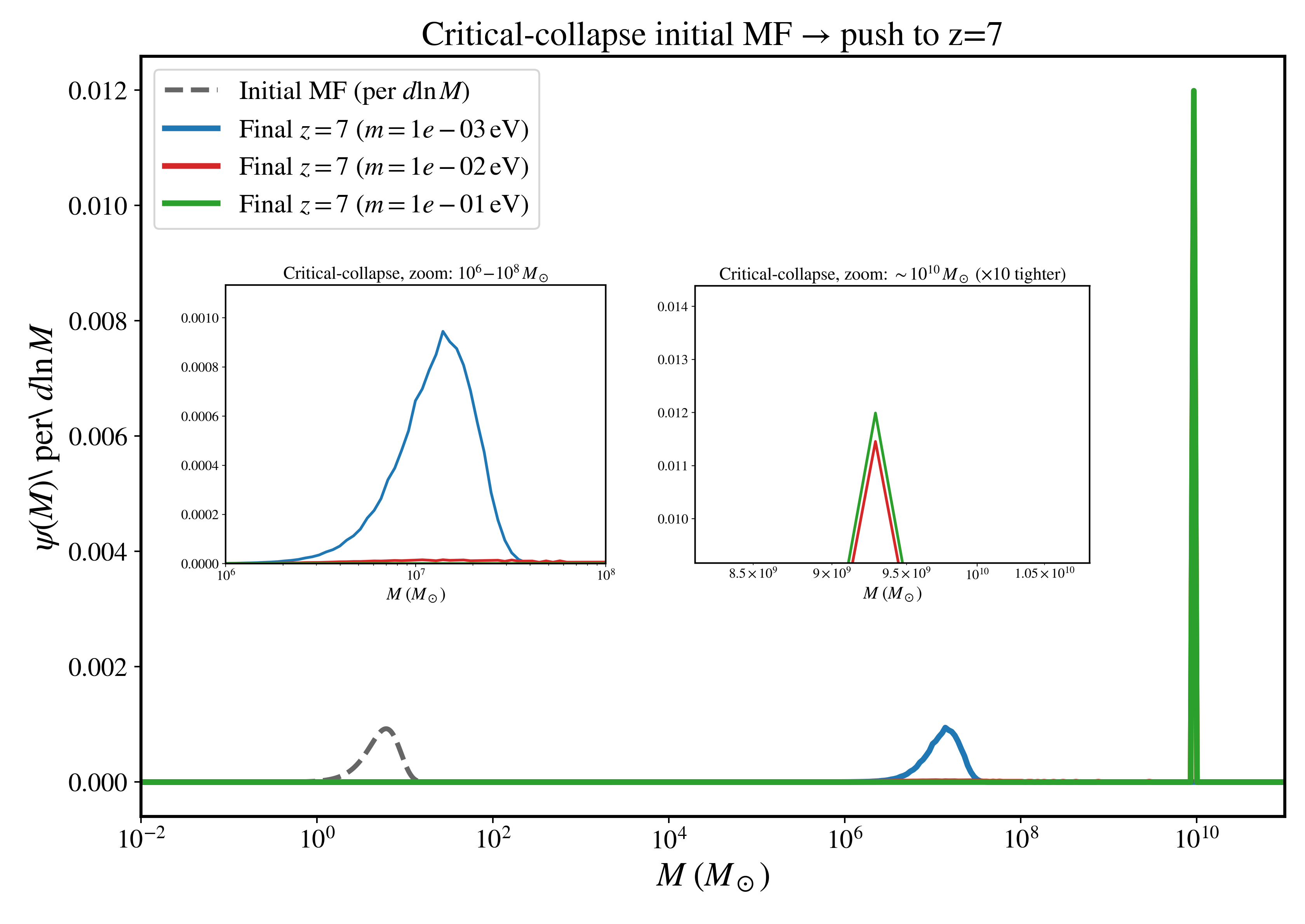}
\caption{Evolution of the QCD PBH mass function for Log-normal(top), power-law (middle), and critical-collapse (down) distributions to $z=7$ under SIDM-Bondi-aided Eddington accretion with DM mass $m=10^{-3,-2,-1} \; {\rm eV}$. The growth triggers early Bondi dominance for $m\gtrsim 10^{-2} \; {\rm eV}$ to reach $10^{10} M_\odot$ SMBHs. The insets in each figure show the finer structures of the final mass function.}
\label{fig:MFz=7}
\end{figure}

We adopt a subdominant PBH abundance $f_{\rm{PBH}}=10^{-3}$, consistent with the limits from CMB anisotropy and spectral-distortion constraints on early energy injection ($f_{\rm{PBH}}\lesssim10^{-3}$)~\cite{Ricotti:2007au,AliHaimoud2017,Poulin:2017bwe,Serpico2020}. Independent bounds from microlensing surveys (MACHO, EROS, OGLE, Subaru/HSC) likewise rule out $M\sim10^{-10}-10^2M_{\odot}$ PBHs as the dominant DM component~\cite{Tisserand2007,Niikura2019,CarrKuhnel2020}. Although QCD PBHs cannot supply the total dark matter, subdominant 
populations remain viable and can participate in astrophysical growth, providing a physically motivated population of early seeds.

%PBHs sourced at the QCD epoch are commonly described by a log-normal  distribution,
%\begin{equation}
%\psi_0(M_0)=\frac{f_{\rm PBH}}{\sqrt{2\pi}\sigma M_0}
%\exp\!\left[-\frac{(\ln(M_0/M_c))^2}{2\sigma^2}\right],
%\end{equation}
%here we choose the central mass $M_c = 5\,M_\odot$ and width $\sigma = 0.5$, reflecting the characteristic horizon mass at $T\sim 200$\,MeV. %Figure~\ref{fig:PBH-QCD_exclusion} shows the resulting spectrum. 

%We adopt a subdominant PBH abundance $f_{\rm{PBH}}=10^{-3}$, consistent with limits from CMB anisotropy and spectral-distortion constraints on early energy injection ($f_{\rm{PBH}}\lesssim10^{-3}$)~\cite{Ricotti:2007au,AliHaimoud2017,Serpico2020}.Independent bounds from microlensing surveys (MACHO, EROS, OGLE, Subaru/HSC) likewise rule out $M\sim10^{-10}-10^2M_{\odot}$ PBHs as the dominant DM component~\cite{Tisserand2007,Niikura2019}. Although QCD PBHs cannot supply the total dark matter, subdominant populations remain viable and can participate in astrophysical growth, providing a physically motivated population of early seeds.

%\begin{figure}[t]
%\centering
%\includegraphics[width=0.48\textwidth]{graphs/PBH-QCD_exclusion.png}
%\caption{
%Log-normal PBH mass function expected from QCD-induced collapse, with $M_c=5\,M_\odot$ and variance $\sigma=0.5$.  Observational constraints limit the total abundance to $f_{\rm PBH}=10^{-3}$.}
%\label{fig:PBHQCD_exclusion}
%\end{figure}

%\subsection{Mass function at $z=7$ and the probability of forming SMBHs}

Using the analytical mass-growth solution, we propagate the full initial PBH distribution forward to $z=7$. If we ignore the PBH merger events, by imposing number conservation, $\psi_{f}(M,z)dM=\psi_{0}(M_{0})dM_{0}$, gives the evolved mass function at any redshift : 
\begin{equation}
    \psi_{f}(M,z)=\frac{\psi_0(M_{0}(M,z))e^{\Gamma\tau{z}}}{[e^{\Gamma\tau{z}} + \frac{K(m)}{\Gamma}M(z)(e^{\Gamma\tau{z}} -1)]^2},  
\end{equation}
where $\psi_0(M_0(M,z))$ refers to each initial function with $M_{0}(M,z)$ obtained by the inverse of \eq{eq:MassGrowthClosedFormMain}. 
%\begin{equation}
%    M_{0}(M,z)=\frac{M(z)}{e^{\Gamma t} + \frac{K(m)}{\Gamma}M(z)(e^{\Gamma t} -1)}.
%\end{equation}
We plot the evolved mass functions at $z=7$ in Fig. \ref{fig:MFz=7} for the above three PBH mass functions and two representative values of scalar mass, $m=10^{-3,-2,-1}\; {\rm eV}$. The results tell some interesting features. Once the SIDM mass exceeds $m \gtrsim 10^{-2}\,\mathrm{eV}$, PBHs enter the Bondi–dominated phase early, leaving ample cosmic time for nearly the entire initial population to exhaust the halo core. It results in a narrow spike around $10^{10}\,M_\odot$ in the final mass function for all three PBH mass–function prescriptions. Moreover, the power–law case is particularly revealing: its extended low–mass tail stalls near $10^{6}\,M_\odot$ with almost the same form in the final mass function. However, the other two cases leave no trace of their initial forms. In contrast, the final mass functions almost maintain their initial PBH forms but shift to a higher mass scale if $m\lesssim 10^{-2} \; {\rm eV}$.

\section{Conclusion}

%\vspace{-1.5cm}
%\medskip
%\noindent\textbf{Conclusion—}
We have demonstrated that PBHs formed during the QCD epoch can evolve into the supermassive regime when embedded in a relativistic SIDM fluid. In the minimal $\phi^4$ model studied here, the entire growth history-and thus the full mapping from the initial PBH distribution to the mass function at the cosmic dawn-is fixed by a single microscopic parameter, the SIDM mass $m$. This yields a predictive, falsifiable mechanism for early SMBH formation, motivating the extension of this microphysical framework to a broader class of SIDM models. 

%\bigskip 

%\noindent {\it Acknowledgements.} 
\begin{acknowledgments}
%\vspace{-1.5cm}
The works of CSC and FLL are supported by Taiwan's National Science and Technology Council (NSTC) through Grant No.-112-2112-M-032 -012 - and No.~112-2112-M-003-006-MY3, respectively.  
\end{acknowledgments}

%\pagebreak

\begingroup
\renewcommand{\addcontentsline}[3]{}% Remove functionality of \addcontentsline
\renewcommand{\section}[2]{}% Remove functionality of \section
\section{References}

\bibliographystyle{apsrev4-2}
\bibliography{refs}   % your references
\endgroup

\clearpage
%%%%%%%%%% Merge with supplemental materials %%%%%%%%%%
\widetext
\setcounter{section}{0}
\setcounter{equation}{0}
\setcounter{figure}{0}
\setcounter{table}{0}
\setcounter{page}{1}
\setcounter{tocdepth}{2}
\makeatletter
\renewcommand{\theequation}{S\arabic{equation}}
\renewcommand{\thefigure}{S\arabic{figure}}
\renewcommand{\bibnumfmt}[1]{[S#1]}
\renewcommand{\citenumfont}[1]{#1}

%%%%%%%%%% Prefix a "S" to all equations, figures, tables and reset the counter %%%%%%%%%%
\begin{center}
\textbf{\large Supplementary Material for \textit{A Unified Dark-Matter–Driven Relativistic Bondi Route to Black-Hole Growth from Stellar to Supermassive Scales}}
\end{center}

%\tableofcontents

\bigskip
\bigskip
\bigskip

In this supplementary material, we will first discuss the possibility of forming super-Eddington supermassive black holes (SMBHs), meaning black holes with masses $\ge 10^9 M_\odot$ formed at redshift $z\simeq 6\textrm{--}7$. Then, we provide full details on solving the Bondi accretion for $\phi^4$-SIDM and use the critical regime to obtain the results used in the main text. We end with two remarks: (i) dark halos as TOV configurations of SIDM; (ii) critical Bondi accretion of $\phi^n$-SIDM. Besides, we also address the Goldilocks problem of $\phi^4$-SIDM and the possible resolutions. 

\section{Forming super-Eddington SMBHs from Bondi accretion?}

The two typical accretion mechanisms for the formation of massive black holes are the Eddington accretion and Bondi accretion. We will discuss whether the Bondi-aided Eddington accretion can produce the super-Eddington SMBHs.

The accretion rate of Eddington accretion is given by
\be\label{Edd_acc}
\dot{M}_{\rm Edd} = {L_{\rm Edd} \over \eta c^2}\;, \quad L_{\rm Edd} = {4 \pi G M m_p c \over \sigma_T}
\ee 
where $M$ is the black hole mass, $\eta$ is the radiative efficiency (taken to be $\eta=0.1$ in the fiducial case), $L_{\rm Edd}$ is the Eddington luminosity expressed in terms of Thompson cross section $\sigma_T$ and proton mass $m_p$, etc. From \eq{Edd_acc}, one can find $M(t) \sim e^{t/t_{\rm Edd}}$ with the characteristic time scale $t_{\rm Edd}\simeq 4.5 \times 10^8$ yrs.

On the other hand, the Bondi accretion rate is given by \citep{Bondi1952, HoyleLyttleton1941, Shapiro:1983du}
\be\label{Bondi_acc_0}
\dot{M}_{\rm Bondi} = 16 \pi \gamma(\Gamma) \rho  {(G M)^2 \over a^3 } 
\ee
for the fluid with an EoS denoted by $\Gamma$ and with some $\Gamma$-dependent adiabatic index $\gamma$. If the fluid is non-relativistic, then the sound speed $a$ is much less than the speed of light $c$. On the other hand, for a relativistic fluid, $a$ could be of the order of $c$.

Naively, the Bondi accretion rate appears larger, as it has one power of $M$ higher than the Eddington rate. However, in reality, the Bondi accretion is suppressed in comparison to the Eddington accretion due to the following reasons:
\begin{enumerate}
 
 \item The Eddington accretion is due to the balance between the radiation pressure and gravity, and the heat due to the collision can be depleted by the radiation to prevent the heating-up feedback.

 \item Unlike the Eddington accretion rate with a universal $t_{\rm Edd}$, the Bondi one depends on energy density $\rho$ of the accreting fluid, which is usually insufficient to supply enough inflow of matter to form SMBHs. For examples, for a black hole of $M=3 M_\odot$ yielding $\dot{M}_{\rm Edd}\sim 6 \times 10^{-8} M_\odot/{\rm yr}$, then (i) in the post-ionization intergalactic medium with $n \sim 10^{-6} {\rm cm}^{-3}$, $a \sim 12 {\rm km}/{\rm s}$, 
\be
\dot{M}_{\rm Bondi}\sim 10^{-12} \dot{M}_{\rm Edd}\;,
\ee
and (ii) in the molecular cloud with $n \sim 100 {\rm cm}^{-3}$, $a \sim 0.3 {\rm km}/{\rm s}$,  
\be
\dot{M}_{\rm Bondi}\sim 10^{-3} \dot{M}_{\rm Edd}\;.
\ee

\item Moreover, considering the accretion of cold dark matter (CDM), the local density of CDM in the inner region is estimated to be far smaller than the baryonic matter in the halos with NFW structures \citep{Blumenthal1986AdiabaticContraction, Navarro1996NFW, Mo2010GalaxyFormation, Bryan2013EAGLE}.   This leads to a suppressed accretion rate compared to Bondi's prediction.

 \item On the other hand, considering the accretion of relativistic dark matter, the Bondi accretion rate is suppressed by large $c_s \sim {\cal O}(c)$. This can also be understood as resistance to collapse under high pressure, leading to slower accretion. 

\end{enumerate}

Therefore, in the usual consideration of forming SMBH by accretion mechanism, the Eddington is usually the dominant mechanism, and the contribution from Bondi accretion can be neglected.  However, before moving into the discussion of bypassing the above no-go of forming SMBH by Bondi accretion, we shall discuss the seeming divergence of Bondi accretion rate of \eq{Bondi_acc_0} for $a=0$. 

\begin{enumerate}

\item The fact of zero sound speed, $a=0$, implies no pressure support, thus no Bondi flow. Thus, the $a \rightarrow 0$ is a critical limit, so that the Bondi accretion should be replaced by ballistic free fall of dust with the accretion rate
\be\label{Mdot_bal}
\dot{M}_{\rm dust} = 4 \pi r_{\rm cap}^2 \rho_{\infty} v_{\infty} = 16 \pi \rho_{\infty}  {(G M)^2 \over v_{\infty}^3 }
 \ee
where $r_{\rm cap}={2 G M \over v^2_{\infty}}$ is the gravitational capture radius, and $v_{\infty}$ is the asymptotic ballistic velocity. Similarly, $v_{\infty}\rightarrow 0$ is also a critical limit of \eq{Mdot_bal} because the configuration becomes static.

\item Physically, the above failure of Bondi accretion in the critical limit can be understood as follows. The vanishing $a$ implies the CDM is almost collisionless, so that its capture by the central black hole simply relies on gravity. This needs a small impact parameter, resulting in a smaller effective accretion rate than predicted by Bondi accretion.

 \item The tiny scattering cross-section of CDM also limits the increase of coarse-grained phase-space density to form ''Bondi radius". Thus, it lacks a mechanism to realize huge "Bondi focusing" even with a tiny sound speed. 

\end{enumerate}

We now move the consider how to bypass the no-go of forming SMBH by Bondi accretion. From the above discussions, we cannot consider CDM, but need to 
pick up a relativistic DM model which can result in a large prefactor of the Bondi accretion rate, at least in the late-time era. Additionally, we need to address the supply issue by ensuring a dense profile in the inner region. 

We find that the SIDM made of a light scalar with quartic self-interaction can bypass the above obstructions because
\begin{enumerate}

\item The prefactor of the Bondi accretion rate diverges as $a \rightarrow c/3$.

\item The Bondi accretion creates a spike profile around the center black hole as shown in \citep{Feng2022_SIDM_Bondi}. This ensures a dense profile to supply Bondi accretion. 

\item This SIDM model can also resolve other astrophysical issues, such as the satellite problem, bullet cluster collision, and the too-big-to-fall problem, and satisfy the cosmological constraint by appropriately tuning the model parameters, e.g., see \citep{SpergelSteinhardt2000, Randall2008, Zavala2013SIDM, vandenAarssen:2012vpm, Tulin:2012wi, TulinYu2018}.

\item The Bondi accretion rate should be self-regulated to avoid the over-production of SMBHs in the current era, which we do not observe. 

\item The total mass of the halos can also be estimated by treating the primordial halos as the Tolman-Oppenheimer-Volkoff (TOV) configurations. 

\end{enumerate}

The key message from the above discussions is as follows: Assuming that SMBHs form via Bondi accretion of DM, the recently discovered SMBHs by JWST can place a very tight constraint on DM models.  Once a viable DM model is chosen, it can predict the mass function and growth curves of SMBHs, assuming that accretion is the primary mechanism for SMBH formation. Future observations can then test these mass functions and growth curves.  

\section{Bondi accretion of $\phi^4$-SIDM}

In this section, we discuss the Bondi accretion of $\phi^4$-SIDM in detail, and the particular regime leading to the environment-independent Bondi accretion rate for the formation of super-Eddington SMBHs. 

\subsection{Relativisitc $\phi^4$-SIDM}
The growth of PBH seeds is modeled within a two-fluid accretion picture, in which baryonic gas accretes under the Eddington limit, with its accretion rate given in \eq{Edd_acc}. 
The DM component is taken to accrete through a Bondi-type process. However, we will not consider the CDM model, but rather the SIDM, which is composed of a relativistic massive scalar field $\phi$ with a quartic self-interaction, i.e.,
\be\label{SIDM_model}
\mathcal{L} \;\supset\; \frac{1}{2}(\partial\phi)^2 - \frac{1}{2}m^2 \phi^2 - \frac{\lambda}{4!}\phi^4 \;.
\ee
We call this model $\phi^4$-SIDM. By considering it under the self-gravitating process in forming a Tolman-Oppenheimer-Volkoff configuration, we can derive the equation of state (EoS) of $\phi^4$-SIDM  \citep{Colpi:1986ye, Zhang:2023hxd}, which takes the following analytic form
\be\label{SIDM_EoS}
 \frac{\rho}{\rho_{B}}=\frac{3p}{\rho_{B}}+4\sqrt{\frac{p}{\rho_{B}}}
\ee
with a model energy density scale  
\be\label{rho_B_0}
 \rho_B := \frac{3m^4}{2\lambda}=\frac{3.48}{\lambda}\Big(\frac{m}{\rm{GeV}}\Big)^{4}\times10^{20} ~ \rm{kg~m^{-3}}.  
\ee
From EoS \eq{SIDM_EoS}, we can obtain the (adiabatic) sound speed square,
\be\label{sound_speed}
a^2 := \Big(\frac{\partial p}{\partial \rho}\Big)_{\rm{adiabatic}}=\frac{1}{3}\Big(1-\frac{1}{\sqrt{1+3\rho/4\rho_{B}}}\Big)\;.
\ee
In the above, we set the speed of light $c$ to unity. From now on, we will adopt the convention of units with $G=c=1$. From \eq{sound_speed}, we note that 
\be
a^2 \le 1/3\;.
\ee
It is called the sound barrier when the bound is saturated.

Besides, for a relativistic fluid, the rest mass density $\rho_0$ is different from the energy density $\rho$, and is related by
\be\label{rho0}
\Big( {\partial \rho \over \partial \rho_0} \Big)_{\rm adiabtic} = {p + \rho \over \rho_0}\;. 
\ee
With EoS \eq{SIDM_EoS}, we can integrate \eq{rho0} to obtain
\be\label{rho02a}
{\rho_0 \over \rho_{0,\infty}} = {a^2 \over a^2_{\infty}}\sqrt{(1- a^2) (1-3 a^2_{\infty})^3 \over (1-a^2_{\infty}) (1- 3 a^2)^3 }  
\ee
where the subscript $\infty$ indicates the quantity being evaluated at spatial infinity. The asymptotic sound speed $a_{\infty}$ also defines the Bondi radius
\be\label{Bondi_radi}
r_B:={2 M \over a^2_{\infty}}\;,
\ee
which characterizes the effective capture radius for the Bondi accretion. For a fixed $a_{\infty}$, we see that the Bondi radius grows linearly with $M$, that is, the matter inside a halo will be gradually accreted into the center black hole as time goes by. 

On the other hand, the $\phi^4$-SIDM model \eq{SIDM_model} gives a cross-section of DM-DM scattering as follows \citep{Eby:2015hsq} 
\be\label{sigma_phi4}
\sigma_{\phi^4} = {\lambda^2 \over 64 \pi m^2}\;.
\ee
The most robust empirical constraints on the cross-section of DM-DM scattering arise from merging clusters, most notably the Bullet Cluster~\citep{Clowe2006,Randall2008}. The persistence of spatial offsets between the hot intracluster gas and DM distribution places an upper bound on the momentum-transfer cross-section per unit mass, 
\begin{equation}
  0.1~\rm{cm^2\,g^{-1}} \lesssim\sigma_{\rm DM}/m \lesssim \mathcal{O}(1)~{\rm cm^2\,g^{-1}}.
  \label{SIDMconstraint}
\end{equation}
This bound has guided SIDM model building, constraining both particle physics scenarios and astrophysical halo structure. Identifying $\sigma_{\rm DM}$ in \eq{SIDMconstraint} as $\sigma_{\phi^4}$ of \eq{sigma_phi4} gives 
\begin{equation}
    30\Big(\frac{m}{\rm{GeV}}\Big)^{3/2} \lesssim \lambda \lesssim 90\Big(\frac{m}{\rm{GeV}}\Big)^{3/2}\;. 
\end{equation}
For simplicity, in the following discussion, we take the median value of $\lambda$ in the above range, namely, $\lambda \simeq 50 (m/\rm{GeV})^{3/2}$. Then, the model energy density scale of \eq{rho_B_0} can be reduced to 
\be \label{rho_B_1}
    \rho_{B}=6.96\times10^{18} \; \Big(\frac{m}{{\rm GeV}}\Big)^{5/2} ~ {\rm kg~m^{-3}} = 2.2 \times 10^{-4} \; \Big(\frac{m}{{\rm eV}}\Big)^{5/2} ~ {\rm kg~m^{-3}}\;.
\ee
Thus, we can tune the mass parameter $m$ of $\phi^4$-SIDM to achieve the desired value of $\rho_B$ by satisfying the most robust constraint \eq{SIDMconstraint} on $\sigma_{DM}$ at the same time.

\subsection{Bondi accretion rate of $\phi^4$-SIDM}

After considering the observational constraint on the $\phi^4$-SIDM to arrive at \eq{rho_B_1}, we now further consider its Bondi accretion around a central black hole. As our SIDM originates from a relativistic light scalar field, the Bondi accretion should be relativistic. The Bondi accretion is governed by the Euler equation \citep{Shapiro:1983du}
\be
\Big( {p + \rho \over \rho_0} \Big)^2 \Big(1+ u^2 -{2 M \over r} \Big) = \Big( {p_{\infty} + \rho_{\infty} \over \rho_{0,\infty} } \Big)^2
\ee
where $u$ is the (inward) stream velocity of the accretion flow, and the accretion rate is given by
\be
\dot{M}_{\rm Bondi} = 4\pi r_s^2 \rho_{0,s} u_s 
\ee
where the subscript $s$ indicates the quantities being evaluated at the sonic horizon $r=r_s$, on which the Euler equation is degenerate and leads to the following critical conditions
\be
u_s^2={a_s^2 \over 1 + 3a_s^2} ={M \over 2 r_s}\;. 
\ee
Using these conditions and the Euler equation, it leads to
\be\label{a2as}
a^2_{\infty} = {a_s^2 (9 a_s^2 -1) \over 2 - 3 a_s^2 + 9 a_s^4}
\ee
which implies
\be
1/9\le a_s^2 \le 1/3\;, \qquad 0\le a_{\infty}^2 \le 1/3\;.
\ee
Using \eq{rho02a} and \eq{a2as}, we can obtain the accretion rate in terms of $a_s$ as follows\footnote{The form is different from the one in \citep{Feng2022_SIDM_Bondi} due to the different choices of boundary condition. Here, choose the boundary condition to make the accretion rate explicitly proportional to $\rho_{0,\infty}$. },
\be\label{accretion-SIDM}
\dot{M}_{\rm Bondi} =2\pi\frac{(1+ 3 a^2_{s})^3}{a^3_{s}(9a^2_{s}-1)}M^2\rho_{0,\infty}\;.
\ee
Unlike the Bondi accretion rate given in \eq{Bondi_acc_0} for the polytropic type of matter with an artifact pole at $a=0$, the accretion rate of  \eq{accretion-SIDM} of $\phi^4$-SIDM is divergent at $a_s =1/3$. As discussed, this divergence at nonzero $a_s$ should be physical, and can bypass the no-go of forming SMBHs via Bondi accretion. For our purposes, we will consider Bondi accretion in this regime. It is more convenient to introduce a small parameter of characteristics,
\be\label{eps_def}
\epsilon := a_s^2 - {1\over 9} \ll 1\;,
\ee
so that 
\be\label{mass_rate_singular}
\dot{M}_{\rm Bondi}={128  \pi \over 9} {1\over \epsilon} M^2\rho_{0,\infty}\;.
\ee
Note that in this critical regime, the sonic horizon radius,
\be
r_s \simeq 24 M \;.
\ee

By the consistency relations of Bondi accretion formalism, we can express $\epsilon$ as follows\footnote{The steps are as follows. First, combining \eq{rho02a} and \eq{sound_speed}, we obtain $\rho_{0,s}={2 (1+3 a_s^2)^{3/2} \rho_{0,\infty} \over 9 a_s^2 -1}$. We further expand the above by using \eq{eps_def} to get $\rho_{0,s}={16 \rho_{0,\infty} \over 27 \sqrt{3} \epsilon} + {\cal O}(\epsilon^0)$, thus second equality of \eq{cons_rel}. Next, using \eq{sound_speed} to solve $\rho_{\infty}$ in terms of $a_{\infty}$, then expand it in order of $\epsilon$ by using \eq{a2as} and \eq{eps_def}, we obtain $\rho_{\infty}={9 \rho_B \epsilon \over 2} + {\cal O}(\epsilon^2)$, thus the first equality of \eq{cons_rel}. Finally, expanding $\rho_{
\infty}$ of \eq{a2as} by using \eq{eps_def} to obtain $a^2_{\infty}={9\epsilon \over 16} + {\cal O}(\epsilon^2)$, thus the third equality of \eq{cons_rel}.}:
\be\label{cons_rel}
\epsilon \simeq {2\over 9}{\rho_{\infty} \over \rho_B} \simeq {16 \over 27\sqrt{3}} {\rho_{0,\infty} \over \rho_{0,s}} \simeq {16 \over 9} a_{\infty}^2\;.
\ee
Plugging the relation given by the third equality of \eq{cons_rel} into the defining \eq{Bondi_radi}, we can obtain the following expression of Bondi radius,
\be\label{Bondi_rad_eps}
r_B \simeq {9 M \over 8}\epsilon^{-1} \;.
\label{rB}
\ee
That is, the Bondi radius is large enough to cover the whole halo core in the critical regime even for the stages of small $M$. This bypasses the no-go of the conventional Bondi accretion in the collisionless limit, and also lifts the supply issue if the halo is heavy enough to form super-Eddington SMBHs. By super-Eddington SMBHs, we refer to the SMBHs with masses larger than $10^9\; M_\odot$, which cannot be produced by the Eddington accretion. 

Furthermore, no relation between $\rho_{\infty}$ and $\rho_{0,\infty}$ can be obtained in the formalism of Bondi accretion. However, one finds that $p_{\infty} \simeq {81 \over 64} \rho_B \epsilon^2$, which is one order of $\epsilon$ more suppressed than $\rho_{\infty} = {9 \over 2} \rho_B \epsilon$, so that it can be considered as pressureless at spatial infinity, thus we have 
\be
\rho_{\infty}\simeq \rho_{0,\infty}:= \rho_{\rm DM}\;, \;\; \because \; p_{\infty} \sim {\cal O}(\epsilon^2)\;.
\ee
Here, we identify $\rho_{0,\infty}$ the DM density inside a typical halos, denoted by $\rho_{\rm DM}$. Moreover, $p_{\infty} \sim {\cal O}(\epsilon^2)$ implies $p_{\infty}\rightarrow 0$. This is a natural boundary condition as the halo is of finite size, beyond which there is no DM. Then, the first relation of \eq{cons_rel} can be replaced by
\be\label{para_char}
\epsilon \simeq {2 \over 9} {\rho_{\rm DM} \over \rho_B}\;.
\ee
using which the Bondi radius given in \eq{Bondi_rad_eps} becomes
\be\label{Bondi_rad_f}
r_B \simeq {81 \over 16}{\rho_B \over \rho_{\rm DM}} M\;.
\ee
We conclude that the Bondi accretion rate \eq{mass_rate_singular} can be simplified to
\be\label{dotM_rhoB}
\dot{M}_{\rm Bondi} = 64 \pi \rho_B {G^2 M^2 \over c^3}\;, 
\ee
where we have recovered the dependence of $G$ and $c$.
This form of Bondi accretion rates says that, in the critical regime, the mass accretion rate does not depend on the halo's DM density, but is only determined by the black hole mass $M$ and the model parameter $\rho_B$. Given relation \eq{rho_B_1} between $\rho_B$ and the scalar mass $m$ after fixing $\lambda$ by the astrophysical and cosmological constraints on DM, we can further express \eq{dotM_rhoB} as follows
\be\label{dotM_m}
\dot{M}_{\rm Bondi} = 1.45 \times 10^{13} \; \Big({M \over M_\odot}\Big)^2 \; \Big(\frac{m}{\rm{GeV}}\Big)^{5/2} \; {M_\odot \over {\rm yr}} \;.
\ee
This relation is one of the key results of this paper. 
It shows that one can constrain the DM model by observations of SMBHs under the assumption that the super-Eddington SMBHs are due to the accretion of some SIDM (currently it is $\phi^4$-SIDM) at the late time epoch. 

Further, by combining \eq{rho_B_1}, \eq{Bondi_rad_eps} and \eq{para_char} and using the typical ambient DM density around the halos
\be
\rho_{\rm DM}= 5 \times10^{-22} \; (1+z)^3 \; \Delta_{\rm vir} \; \rm{kg~m^{-3}}
\ee
where the typical virial overdensity $\Delta_{\rm vir}\simeq 300$ from spherical-collpase simulation \cite{BryanNorman1998,WhiteRees1978,Blumenthal1984}, 
then the Bondi radius of \eq{Bondi_rad_f} can be expressed as
\be
r_B(z) = {81\over 16} {\rho_B \over \rho_{\rm DM}} M = 3.55 \times 10^2 \; (1+z)^{-3} \; \Big( {m \over {\rm eV}}\Big)^{5/2} \Big(\frac{M(z)}{M_{\odot}}\Big) \; \rm{pc}\;.
\ee
To ensure the growth into a super-Eddington SMBH by the current critical Bondi accretion mechanism, the Bondi radius must be larger than the core radius of the halos $\sim 2 - 10$ kpc. That means, for having a SMBH with mass $\sim 10^9 M_\odot$ around $z\simeq 7$ and take the mean value for the core radius about $6$ kpc, then
\be
r_B(z=7) \ge 6\; {\rm kpc} \quad  \Longrightarrow \quad m \ge 9.44 \times 10^{-3} \; {\rm eV} \simeq 10^{-2}  \; {\rm eV}\;. 
\ee

\subsection{Two remarks}
Before considering the growth curve of the SMBH mass function based on the two-fluid accretion scheme, due to Eddington and Bondi accretion, we have two remarks: (i) dark halos as the TOV configurations of the SIDM; (ii) the critical Bondi accretion for $\phi^n$-SIDM.

The first remark is to estimate the mass and size of the DM halos if they are considered as the Tolman-Volkov-Oppenheimer (TOV) configurations. This differs from the usual formation of cold dark matter (CDM) halos by the Press–Schechter (PS) formalism, which assumes the cold dark matter (CDM) model. The core idea of this formalism is to linearly grow the initial Gaussian density perturbations as the universe expands, until they reach the critical sizes for gravitational collapse via thermal virialization. 

As our $\phi^4$-SIDM is relativistic, the PS formalism may not be applicable. Instead, it is quite natural for DM halos to form through gravitational collapse and result in TOV configurations, which are solved from the TOV equations: 
\be\label{TOV}
{d p \over dr}=-(\rho + p) {{\cal M} + 4 \pi r^3 \rho \over r(r-2 {\cal M}) }\;, \qquad {d {\cal M} \over dr}=4\pi r^2 \rho
\ee
with a given EoS $p=f_{\rm EoS}(\rho)$ for some function $f_{\rm EoS}$. Here, $p(r)$, $\rho(r)$, and ${\cal M}(r)$ are the pressure, energy density, and mass profiles, respectively. The radius $R_{\rm TOV}$ of the compact object is determined by $\rho(R)=0$ and its mass $M_{\rm TOV} ={\cal M}(R_{\rm TOV})$. By inspection, the above TOV equations are invariant under the following scaling transformation \cite{PhysRevD.96.023005, Zhang:2023hxd}:
\be\label{scale_1}
&& \rho \rightarrow k_s^{-2} \rho\;, \qquad p \rightarrow k_s^{-2} p\;, 
\\ \label{scale_2}
&& {\cal M} \rightarrow k_s {\cal M} \;,\qquad r\rightarrow k_s r\;.
\ee
From the \eq{SIDM_EoS} of $\phi^4$-SIDM in which $\rho_B$ is adopted as a unit for $\rho$ and $p$, this implies that the TOV configurations and hence their $M_{\rm TOV}-R_{\rm TOV}$ relations can be solved in terms of the dimensionless quantities as long as we also scale $\rho_B$ by
\be
\rho_B \rightarrow k_s^{-2} \rho_B\;. 
\ee
Using \eq{rho_B_1}, this implies that the scalar mass $m$ should scale as
\be
m \rightarrow k_s^{-4/5} m\;,
\ee
if the $M_{\rm TOV}$ and $R_{\rm TOV}$ by a $k_s$ factor.

\begin{figure}
\centering
\includegraphics[width=0.48\textwidth]{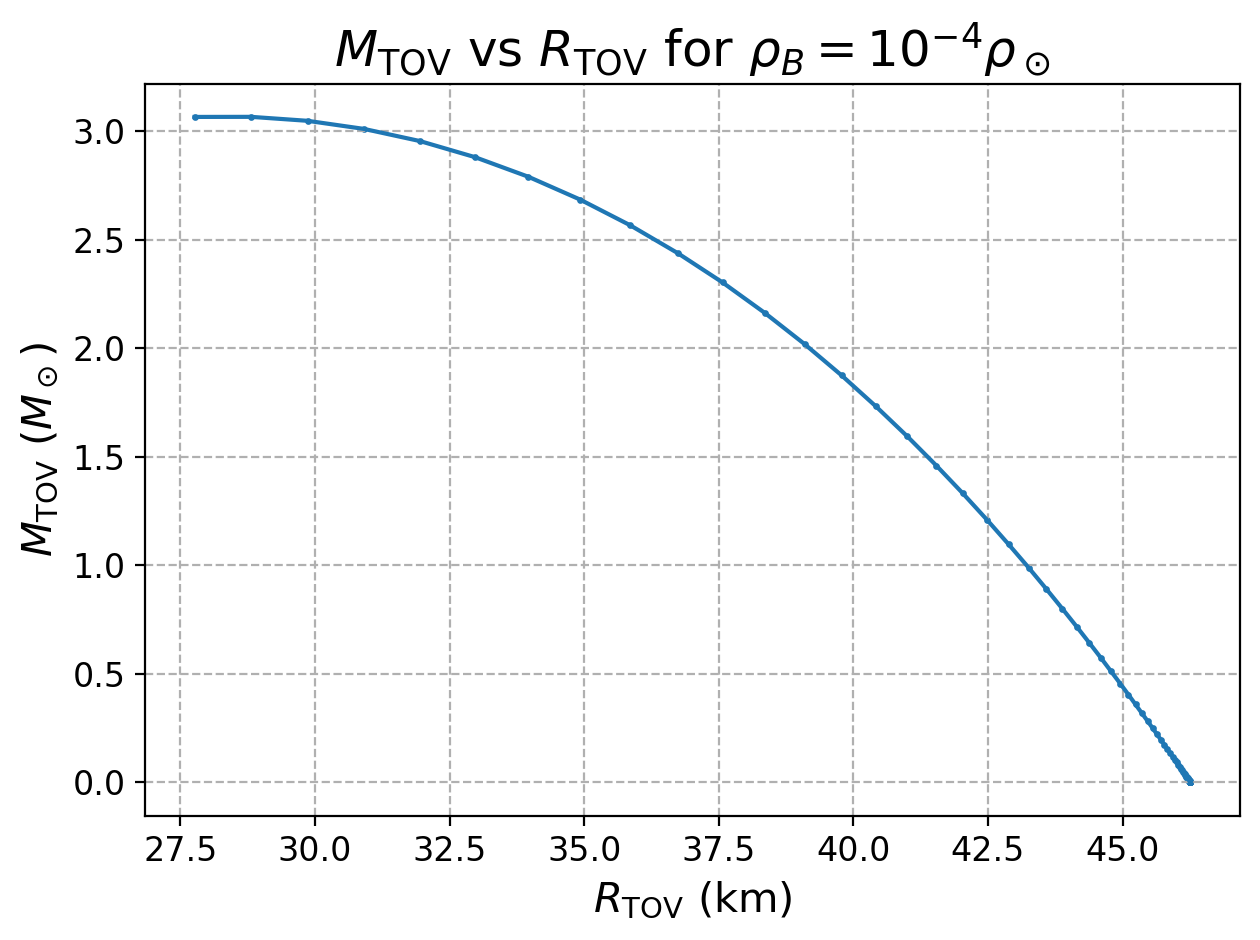}
\caption{Mass-Radius relation for the TOV configurations of $\phi^4$-SIDM with $\rho_B=10^{-4} \rho_\odot$.}
\label{fig:MR}
\end{figure}

In Fig. \ref{fig:MR}, it shows the mass-radius relation for the TOV configurations of $\phi^4$-SIDM with $\rho_B=10^{-4} \rho_\odot$, which corresponds to $m=0.15\; {\rm GeV}$ by using \eq{rho_B_1} \footnote{Compare \eq{SIDM_EoS} and (27) in \cite{Zhang:2023hxd}, we find the relation between $\rho_B$ and ${\cal B}_4$ in \cite{Zhang:2023hxd} as
\be
{\cal B}_4 = 4 \sqrt{\rho_B \over \rho_\odot}\;.
\ee
Using $\rho_\odot=6.18 \times 10^{20} {\rm kg/m^3}$ and \eq{rho_B_0}, we obtain
\be
{\cal B}_4 = {3 \over \sqrt{\lambda}} \Big({m \over {\rm GeV}}\Big)^2 = {3\over \sqrt{50}}\Big({m \over {\rm GeV}} \Big)^{5/4} \;.
\ee
where we have used $\lambda = 50 (m/{\rm GeV})^{3/2}$ to arrive the second equality. For ${\cal B}=0.04$, it yields $m=0.15 \; {\rm GeV}$.
}. Let's pick up the typical configuration with $M_{\rm TOV}=30 M_\odot$ and $R_{\rm TOV}=3 \; {\rm km}$. Then, we have for a $\phi^4$-SIDM model with 
\be
m= 1.5 \times 10^{-8 \alpha -1} \; {\rm GeV}\;,
\ee
which will yield a TOV halo with
\be
M_{\rm TOV}= 3 \times 10^{10\alpha +1} M_\odot 
\ee
with 
\be
R_{\rm TOV}= 3 \times 10^{10\alpha} \; {\rm km} = 9.72 \times 10^{10\alpha-14}\; {\rm pc}\;.
\ee
Also, the Bondi accretion rate \eq{dotM_m} can be re-expressed as 
\be
\dot{M}_{\rm Bondi}= 1.26 \times 10^{13-20\alpha} \Big({M \over M_\odot}\Big)^2 \;  {M_\odot \over {\rm yr}}
\ee

In the main text, we will consider the scalar mass $m\simeq 10^{-2} {\rm eV}$, which corresponds to $\alpha \simeq 1.27$. In this case, we will have
\be
M_{\rm TOV}\simeq 1.5 \times 10^{14} M_\odot\;, \qquad R_{\rm TOV} \simeq 0.46 \; {\rm pc}\;, \qquad \mbox{for } \; m\simeq 10^{-2}{\rm eV}\;. 
\ee
However, we should expect that the real halos will be less compact than those above, and could be comparable with a typical galaxy-scale dark halo with mass $10^{11}- 10^{12} M_\odot$, radius $\sim 200 - 300$ kpc, and the core radius $\sim 2 - 5$ kpc.

The second remark is about whether the $\phi^4$-SIDM is a special DM model to yield nontrivial Bondi accretion for forming super-Eddington SMBHs. The answer could be yes and no. To see this, we can generalize the $\phi^4$ interaction in \eq{SIDM_model} to $\phi^n$ with $n=4,6,7,8,\cdots$. It is straightforward to show that the Bondi accretion of $\phi^n$-SIDM gives the accretion rate in the critical regime as follows:
\be
\dot{M}_{\rm Bondi} \simeq {\cal O}(1) \; \epsilon^{-{2 \over n-2}} \; M^2 \;\rho_{DM}
\ee
with
\be
\epsilon := a_s^2 - {3n -10 \over 3n + 6}\;.
\ee
We see that the Bondi accretion has the critical point at $a_s^2={3n-10 \over 3n + 6}$, but with less critical behavior as $n$ increases. Therefore, $\phi^4$-SIDM is not special as all the $\phi^{n\ge 4}$ models can yield the critical Bondi accretion, which can be exploited to form super-Eddington SMBHs. However, $\phi^4$-SIDM is the most critical one, and this is special, along with its naturalness and simplicity. One can expect more SIDM models to yield physically viable critical Bondi accretion due to the relativistic feature and the nontrivial interactions and resultant EoS. 

\section{Goldilocks problem and its resolution by modifying $\phi^4$-SIDM}\label{appendix_B}

As discussed at the end of the Introduction, the velocity-independent ${\sigma\over m}$ of $\phi^4$-SIDM cannot help resolve the Goldilocks problem: galaxy-scale constraints require higher values of ${\sigma \over m}$, but the cluster scale constraints require lower values. Instead, the Goldilocks issue can be resolved by considering SIDM with velocity-dependent cross-sections, which are large at low velocities (dwarf galaxies) and small at high velocities (clusters). 

One can achieve the required cross-sections by appropriately appending additional ingredients to our SIDM model without disrupting rapid Bondi accretion in the critical regime. The simplest way is to introduce another SIDMs with the required velocity cross-sections to yield a Goldilocks zone. Unlike our $\phi^4$-SIDM, these new SIDMs will have negligible contributions to the Bondi accretion rate in comparison to the critical $\phi^4$-SIDM. 

An alternative option to adding new SIDMs is to introduce a  mediator scalar field $\varphi$ with the cubic interaction 
\be
{\cal L}' = {1\over 2} (\partial \varphi)^2 - m_{\varphi}^2 \varphi^2 - \lambda_3  \varphi \phi^2 
\ee
which can be shown to yield a Goldilocks zone by appropriate tuning $\lambda_3$ and $m_{\varphi}$, e.g., by some resonant self-interaction as indicated in \cite{Tsai:2020vpi}. Assuming $m_{\varphi} \gg m$, we can integrate out $\varphi$ to obtain an effecrtive $\phi^4$ coupling. This then change $\lambda$ to
\be
\lambda'(p) =\lambda + {\cal O}(1) {\lambda_3^2 \over p^2 + m^2_{\varphi}}\;.
\ee
For simplicity, we have written the effective coupling in the momentum space with $p$ denoting the off-shell 4-momentum of  $\varphi$, or 4-momentum exchange of $\phi$ scattering. This is an effective velocity-dependent quartic interaction that can thus be used to resolve the Goldilocks issue, i.e., larger/smaller $\lambda'(p)$ (cross-sections) for smaller/larger $p^2$ (velocities). On the other hand, when deriving the EoS of the $\phi$-field perfect fluid under self-gravity by taking the homogenous limit ($p \rightarrow 0$) in the gradient expansion, e.g., see Appendix A of \cite{Zhang:2023hxd}, we will arrive at the same EoS but replacing $\lambda$ by $\lambda'(p=0)$. However, this replacement will not affect our discussion of the critical Bondi accretion of $\phi^4$-SIDM by appropriately tuning $\lambda_3$ and $m_{\varphi}$.

%\bibliographystyle{JHEP}
%\bibliography{PRL_main.bib}

\end{document}